\documentclass[journal=jacsat,manuscript=article]{achemso}

\usepackage[version=3]{mhchem} 
\usepackage[T1]{fontenc}       
\usepackage{amsmath,amssymb,upgreek}
\usepackage{color}

\author{Yu Song}
\affiliation[MOE Key Laboratory of Space Applied Physics and Chemistry, and Shaanxi Key Laboratory of Optical Information Technology, School of Science, Northwestern Polytechnical University, Xi'an 710072, China]
{School of Science, Northwestern Polytechnical University}
\author{Siqi Hu}
\affiliation[MOE Key Laboratory of Space Applied Physics and Chemistry, and Shaanxi Key Laboratory of Optical Information Technology, School of Science, Northwestern Polytechnical University, Xi'an 710072, China]
{School of Science, Northwestern Polytechnical University}
\author{Miao-Ling Lin}
\affiliation[State Key Laboratory of Superlattices and Microstructures, Institute of Semiconductors, Chinese Academy of Sciences, Beijing
100083, China]
{ Institute of Semiconductors, Chinese Academy of Sciences}
\author{Xuetao Gan}
\email{xuetaogan@nwpu.edu.cn}
\affiliation[MOE Key Laboratory of Space Applied Physics and Chemistry, and Shaanxi Key Laboratory of Optical Information Technology, School of Science, Northwestern Polytechnical University, Xi'an 710072, China]
{School of Science, Northwestern Polytechnical University}
\author{Ping-Heng Tan}
\affiliation[State Key Laboratory of Superlattices and Microstructures, Institute of Semiconductors, Chinese Academy of Sciences, Beijing
100083, China]
{ Institute of Semiconductors, Chinese Academy of Sciences}
\author{Jianlin Zhao}
\email{jlzhao@nwpu.edu.cn}
\affiliation[MOE Key Laboratory of Space Applied Physics and Chemistry, and Shaanxi Key Laboratory of Optical Information Technology, School of Science, Northwestern Polytechnical University, Xi'an 710072, China]
{School of Science, Northwestern Polytechnical University}

\title[An \textsf{achemso} demo]
  {Extraordinary second harmonic generation in  ReS$_2$ atomic crystals}


\begin{document}



\begin{abstract}

We report the observations of unexpected layer-dependent, strong, and anisotropic second harmonic generations (SHGs) in atomically thin ReS$_2$. Appreciable (negligible ) SHGs are obtained from even (odd) numbers of ReS$_2$ layers, which is opposite to the layer-dependence of SHGs in group VI transition metal dichalcogenides, such as MoS$_2$ and WS$_2$. The results are analyzed from ReS$_2$'s crystal structure, implying second harmonic polarizations generated from the interlayer coupling. Pumped by a telecom-band laser, SHG from the bilayer ReS$_2$ is almost one order of magnitude larger than that from the monolayer WS$_2$. The estimated second-order nonlinear susceptibility of 900 pm/V is remarkably high among those reported in two-dimensional materials.  The laser polarization dependence of ReS$_2$'s SHG is strongly anisotropic and  indicates its distorted lattice structure with more unequal and non-zero second-order susceptibility elements. 

\textbf{keywords:}~second harmonic generation,~transition metal dichalcogenides, ~rhenium disulfide,~ polarization,~ interlayer coupling,~ exciton resonance
\end{abstract}


Atomically thin transition metal dichalcogenides (TMDCs) as one of the most important two-dimensional (2D) materials have drawn more and more attentions because of their intriguing physical and chemical properties resulting from structural symmetry and interlayer coupling~\cite{Fai2013,Shan2016,Xu2017}. High-performance 2D TMDC-based devices have been reported, including high-mobility field effect transistors~\cite{LiaoAM,Xinran}, ultrasensitive and broadband photodetectors~\cite{JianluAM}, and low-threshold monolayer lasers~\cite{Xiaodong,Yuye}, etc. Aside from the most popular TMDCs of group VI materials, e.g., MoX$_2$ and WX$_2$, ReX$_2$ (X denotes Se or S) has recently attracted considerable interests due to their wholly unusual properties. First of all, unlike group VI TMDCs, ReX$_2$ grows in a distorted CdCl$_2$ structure and has strong anisotropy in the chemical bonds~\cite{Friemelt,Lin,Hone}, which consequently leads to the anisotropies of in-plane phonon, electrical, and optical properties~\cite{Etienne,MiaoNC}. For instance, Miao $et~al.$ reported the ratio of carrier mobilities along two principal axes of the few-layer ReS$_2$ exceeds 3. This crystallographic and concomitant electrical anisotropy were employed to construct good performance integrated digital inverters~\cite{MiaoNC}. From a transient optical absorption measurement of a monolayer ReS$_2$, it is revealed the absorption coefficient and transient absorption are both anisotropic for light polarized parallel and perpendicular to the Re atomic chains~\cite{zhaohui}. The anisotropic Raman spectra ~\cite{Wolverson,Weiji} and linearly polarized excitons are revealed in few-layer ReX$_2$ as well~\cite{Heinz,Arora}.  Due to the intrinsically linear dichroism originated from the high in-plane anisotropy in few-layer ReSe$_2$, a polarization sensitive photodetection was implemented successfully~\cite{FaxianACS}.  

Second, the Peierls distortion in the 1T$^\prime$  structure of ReX$_2$~ prevents its ordered stacking and reduces the interlayer couplings, which have been demonstrated to be weaker than those in layered MoS$_2$~\cite{Weiji}. In addition, few-layer ReX$_2$~ behaves similar as monolayer ReX$_2$~, including the direct bandgap, and slightly varied photoluminescence spectra with increased layer numbers~\cite{Tongay}. This is in contrast to the group VI TMDCs, where crossover from indirect to direct bandgap occurs with decreasing the number of layers from bulk to monolayer. As a result, ReS$_2$-based photodetectors exhibit gate-tunable photoresponsivities competitive to those reported in  photodetectors based on graphene, MoSe$_2$, GaS, and GaSe~\cite{FaxianAFM}.

Here, we reveal atomically thin ReS$_2$ has unique second harmonic generation (SHG), representing as the third distinct attribute from that in group VI TMDCs. It was reported that finite slices of group VI TMDCs  with odd-layer thicknesses could yield moderately strong SHG due to the broken inversion symmetry. However,  flakes with even numbers of layers are absent of SHG~\cite{Malard2013,Li2013}. By examining SHGs in ReS$_2$ flakes with different layer-numbers, we find their SHGs exhibit an opposite dependence on the layer number, i.e., even (odd) numbers of ReS$_2$ layers have strong (negligible) SHG. By analyzing the crystal structure, the layer-dependent inversion symmetry broken is revealed. In addition, we find the SH signal from a bilayer ReS$_2$ is about one order of magnitude stronger than that from a monolayer WS$_2$ when pumped by a telecom-band laser. A high second-order nonlinear susceptibility in bilayer ReS$_2$ is evaluated exceeding 900 pm/V. The laser polarization dependences of the SHG in ReS$_2$ are discussed as well, showing a distinct anisotropy determined by the triclinic symmetry as well as a nonlinear susceptibility tensor with more non-zero elements. While the weak interlayer coupling in ReS$_2$ results in slight layer-dependences of Raman spectra, direct bandgap, and photoluminescence, our results unveil SHG in ReS$_2$ could strongly indicate the even or odd numbers of layers. 

We implement the SHG measurements in a reflection geometry with a normal incidence excitation. A fiber-based pulsed laser is chosen as the fundamental pump radiation, which is centered around 1558 nm and has a repetition rate of 18.5 MHz and a pulse width of 8.8 ps. The pulsed laser is focused by an 50$\times$ objective lens with a numerical aperture of 0.75 into a spot size of about 2 $\upmu$m on the sample. The SH signal scattered from the ReS$_2$ sample is collected by the same objective lens, which is then examined by a spectrometer mounted with a cooled silicon CCD. The reflected pump laser is filtered out from the SH signal by a dichroic mirror. 

\begin{figure}[th!]\centering
\includegraphics[width=4.5in]{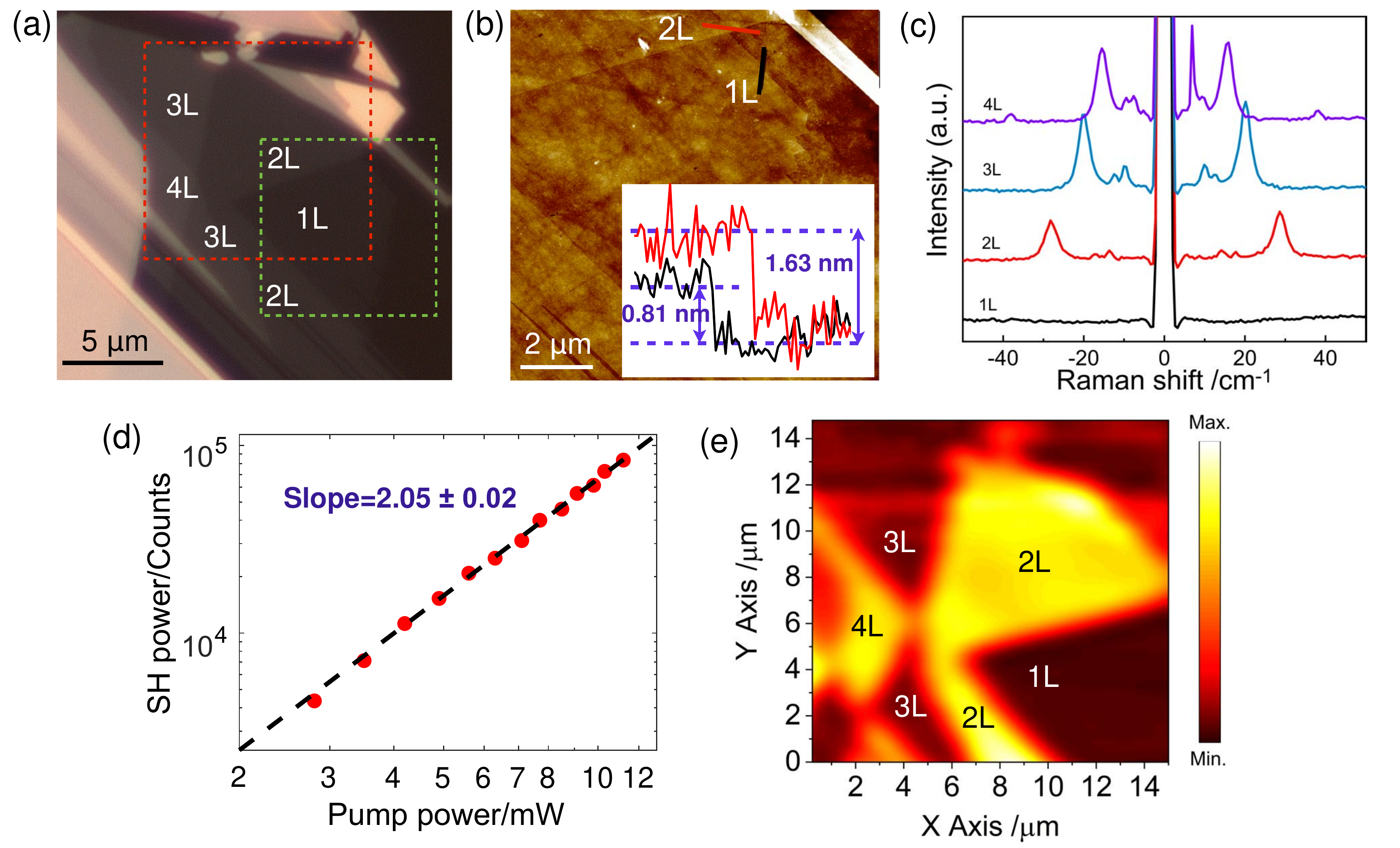}
\caption{{\small  (a) Optical microscope image of the atomically thin ReS$_2$ sample exfoliated on a fused silica. (b) AFM image of the monolayer and bilayer ReS$_2$ crystals, which is indicated by the green dashed box in (a). The inset shows the thicknesses measured along the red and black solid lines. (c) Raman spectra from the ReS$_2$ flakes with different layer numbers.  (d) Log vs log plot of the pump power dependence of SHG from the bilayer ReS$_2$. (e) SHG spatial mapping image of the ReS$_2$ area marked by the red dashed box in (a). }}
\label{fig:imag} 
\end{figure}

Atomically thin ReS$_2$ flakes are mechanically exfoliated on a fused silica or on a silicon covered with 300 nm silicon oxide. Figure~\ref{fig:imag}(a) displays an optical microscope image of a sample with different numbers of layers on a fused silica, showing discernable regions with varied optical contrasts. The thick flakes are easy to break along a well-defined cleavage axis defined by the strong Re-Re atomic bond~\cite{MiaoNC,Heinz}.  The layer numbers are evaluated by atomic force microscopy (AFM). Figure~\ref{fig:imag}(b) shows the measured AFM image of a region marked by the green dashed box in  Figure~\ref{fig:imag}(a). The inset shows the height data measured along the black and red solid lines with values of $\sim$0.81 nm and $\sim$1.63 nm, corresponding to the thicknesses of monolayer and bilayer ReS$_2$, respectively. Raman spectroscopy is a very accurate and efficient technique to distinguish layer numbers of 2D materials by measuring the ultralow-frequency rigid-layer shear modes, which is extremely sensitive to the interlayer coupling~\cite{Weiji,Tan2011,Liang2017}. Figure~\ref{fig:imag}(c) displays Raman spectra acquired from the different layers indicated in Figure~\ref{fig:imag}(a). In the Raman spectrum of the monolayer ReS$_2$, there is no ultralow-frequency Raman mode. The bilayer ReS$_2$ sample is confirmed from the emerging shear and layer-breathing modes at 13.8, 17.5, and 28.7 cm$^{-1}$. And the two shear modes at 13.8 and 17.5 cm$^{-1}$ indicate the anisotropic stacking-order of the two  ReS$_2$ monolayers~\cite{Weiji}. The anisotropic trilayer and teralayer ReS$_2$ samples are recognized further by the numbers and peak positions of the rigid layer modes~\cite{Weiji}. 

In SHG measurements, we first focus the pump laser  on the monolayer ReS$_2$ region, and no any SH signal is detected. When the pump laser illuminates on the bilayer region, however, a strong frequency upconversion signal is obtained around the wavelength of 779 nm. To verify the SH character of the detected radiation, we measure its power dependence on the pump power, as shown in the log-log plot of  Figure~\ref{fig:imag}(d). A nearly quadratic dependence on the pump power with a slope of 2.05 $\pm$ 0.02 is obtained, implying the second-order nonlinear process. Here, limited by the experiment system, the maximum pump power illuminated on the sample is about 11 mW. There is no observable saturation of the SH intensity at this power level. With even higher pump power, deviations from this second-order dependence are expected due to the material damage or electronic population pumping.~\cite{Le2016,Zhang2015} This result is totally different from those presented in group VI TMDCs, which has strong (zero) SHG in monolayer (bilayer) flake~\cite{Malard2013,Li2013}. A SHG spatial mapping is then carried out over the ReS$_2$ region marked in the red dashed box in  Figure~\ref{fig:imag}(a), including mono-, bi-, tri-, and tetra-layers. Figure~\ref{fig:imag}(e) presents the SHG spatial mapping image, showing well-defined regions with varied SH intensities. Strong SH signals are observed over the bilayer and tetralayer regions, and the monolayer and trilayer flakes show zero or negligible SHG.  

\begin{figure}[th!]\centering
\includegraphics[width=4.5in]{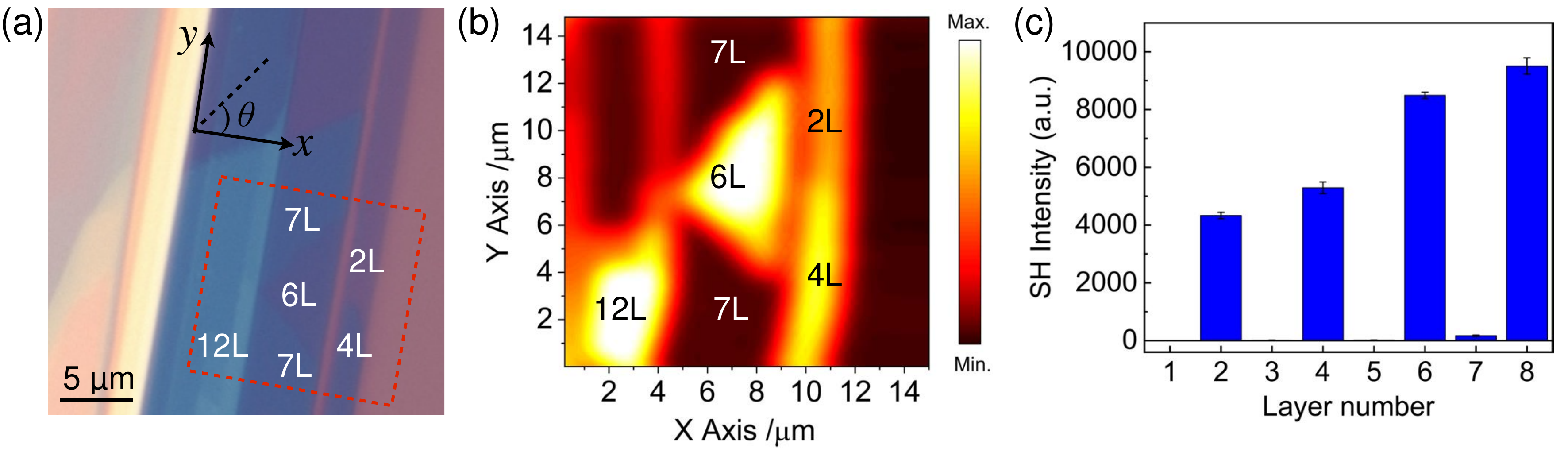}
\caption{{\small  (a) Optical microscope image of a ReS$_2$ sample exfoliated on a silicon covered with 300 nm silicon oxide. (b)  SHG spatial mapping image of the ReS$_2$ flake marked by the red dashed box in (a). (c) Experimentally measured layer dependence of ReS$_2$'s SH intensities, where the uncertainties are indicated by the error bars. }}
\label{fig:layer} 
\end{figure}

This layer-dependent SHG is confirmed further from another ReS$_2$ sample exfoliated on a silicon covered with 300 nm silicon oxide. The optical microscope image of the sample is displayed in  Figure~\ref{fig:layer}(a). Examined by AFM, it has multiple thick layers, as noted in the optical microscope image. A SHG spatial mapping is implemented over the region marked by the red dashed box in  Figure~\ref{fig:layer}(a), as shown in  Figure~\ref{fig:layer}(b). The result agrees well with the conclusion revealed in  Figure~\ref{fig:imag}, showing strong (negligible) SH signals from the even (odd) numbers of ReS$_2$ layers.  To conclude, we plot the layer dependences of ReS$_2$'s SH intensities in  Figure~\ref{fig:layer}(c), presenting sharp SHG contrast between even and odd numbers of layers. SHG in monolayer ReS$_2$ is absolutely zero, and SHG in other odd layers are no more than 0.5\% of those in even layers. The high suppression of SHG in odd numbers of ReS$_2$ layers results from the existence of inversion symmetry, as expected from the bulk ReS$_2$ crystal. We also measure the SH signal from the bulk material and find it is negligible. Strong SHG in even numbers of ReS$_2$ layers indicate the broken inversion symmetry in their crystal structures. 

\begin{figure}[th!]\centering
\includegraphics[width=4.5in]{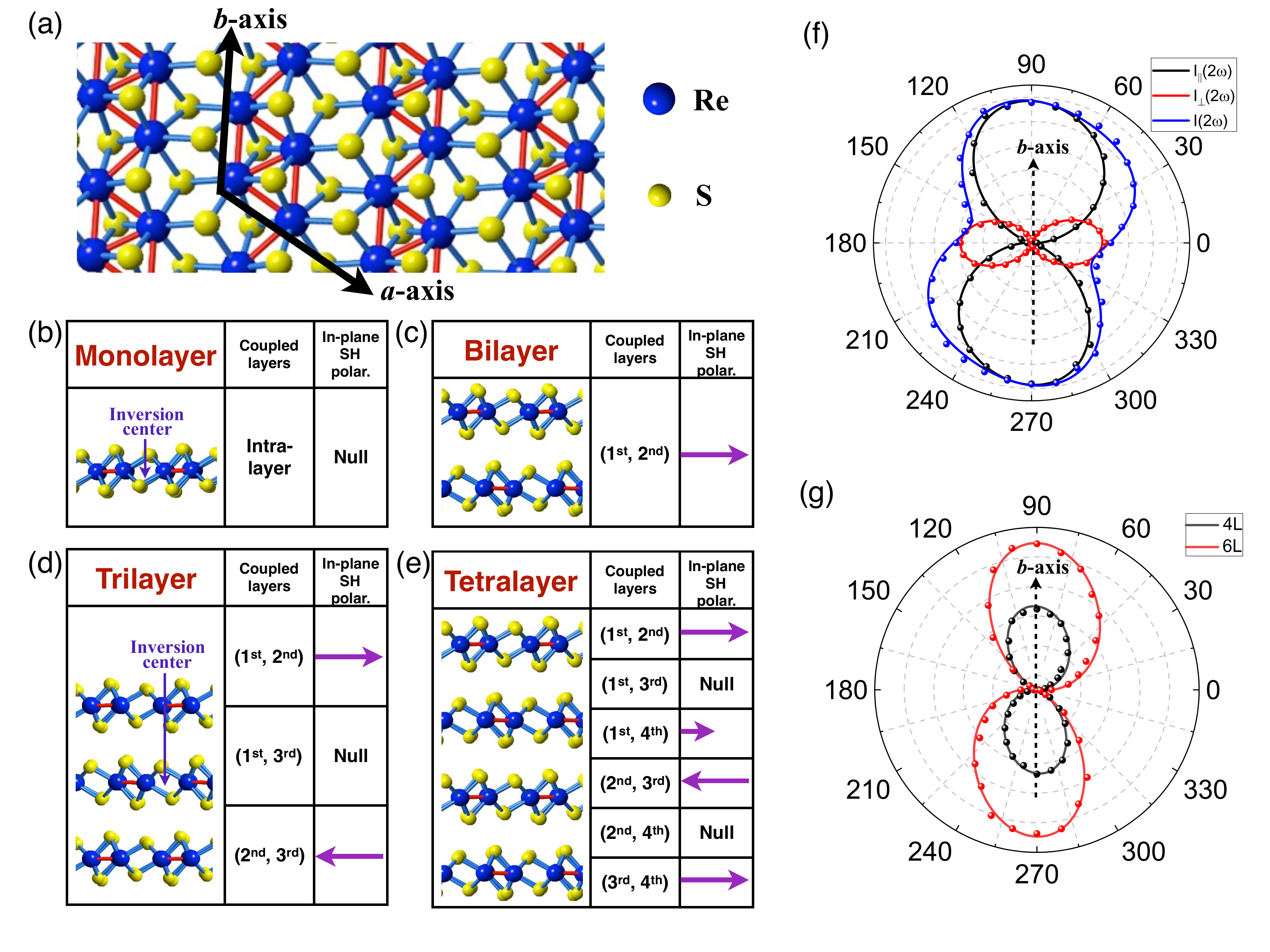}
\caption{{\small  (a) Top view of the crystal structure of monolayer ReS$_2$. (b)-(e) Side views of crystal structures from mono- to tetralayer ReS$_2$. The model for explaining the relation between SH intensities and layer numbers is illustrated as well. In-plane SH polarization is assumed to be generated from the interlayer coupling. (f) Polar plots of the SH intensities from the six-layer ReS$_2$, showing SH radiation components detected parallelly and perpendicularly to the polarization of the pump laser as well as their summation. (g)  Polar plots of parallel components of SH intensities from the tetralayer and six-layer ReS$_2$.}}
\label{fig:stru} 
\end{figure}

To understand the  absence (presence) of inversion symmetry in  even (odd) numbers of ReS$_2$ layers, we draw the crystal structures from mono- to tetra-layer ReS$_2$ in  Figures~\ref{fig:stru}(a)-(e). In monolayer ReS$_2$, each Re atom has six neighboring S sites and is sandwiched by the S atoms at both sides. Figure~\ref{fig:stru}(a) shows the top view of a monolayer crystal structure. Different from group VI TMDCs, four adjacent Re atoms are bonded together to form a zigzag chain, presenting a distorted octahedral lattice structure. The rhombus formed by the four adjacent bonded Re atoms defines two principle axes, the $b$ and $a$ axes, which have an angle of 118.97$^\circ$. The strong Re chain oriented along the direction of lattice vector $b$-axis endows ReS$_2$'s strong anisotropies along the two principle axes, which also reduces its crystal symmetry to the lowest one among the TMDCs. When multiple ReS$_2$ monolayers stack into multilayer ReS$_2$ crystals, there are a variety of possible stacking orders. As demonstrated in Ref. 16, the most stable stacking configuration is that Re-Re bonds in the adjacent layers have an oriented angle of 60$^\circ$, which forms an anisotropic-like multilayer ReS$_2$.  Figures~\ref{fig:stru}(b)-(e) display the side views of the crystal structures from mono- to tetra-layer ReS$_2$.  In monolayer ReS$_2$, there is only one symmetry operation, i.e., the inversion symmetry $i$. The inversion center is indicated by the arrow.  The same symmetry operation holds for other odd numbers of layers, as well as a bulk crystal. ReS$_2$ crystals with this low crystal symmetry belong to the $C_i$ space group.  In even numbers of ReS$_2$ layers, the symmetry operation is reduced further with the broken inversion symmetry, as illustrated in the bilayer and tetralayer crystal structures. Therefore, the even numbers of layers belong to the noncentrosymmetric $C_1$ space group. 

In Figures~\ref{fig:stru}(b)-(e), combining with the  transformations of crystal symmetry for different layer-numbers, we also illustrate a simple model to analyze the layer-dependent SH intensities. The inversion symmetry in monolayer ReS$_2$ makes its SHG to be exactly zero. In bilayer ReS$_2$, each layer can not generate SH signal either due to the intralayer inversion symmetry. Hence, the observed strong SH radiation in bilayer ReS$_2$ could be attributed to the appreciable SH polarization generated from the interlayer coupling. It is consistent with the inversion symmetry broken when the two monolayers stack into a crystal structure shown in  Figure~\ref{fig:stru}(c). The crystal structure also indicates that the interlayer SH polarization should result from both the in-plane and out-of-plane dipoles. Considering the low numerical aperture of the employed objective lens, our experimentally measured SH signal mainly arises from the in-plane SH polarization. We therefore use the in-plane SH polarization  to qualitatively explain the layer-dependent SHG, as shown in Figures~\ref{fig:stru}(b)-(e). By using a tilted far-field microscopy,{~\cite{Janus1} near-field plasmonic modes{~\cite{Markus1,Park1}, or manifold field-gradients in a surface-plasmon,~\cite{Markus2} the out-of-plane SH polarizations could be further examined in future work. With the measured in-plane and out-of-plane SH polarizations, a complete model of SH polarizations induced by the interlayer coupling could be constructed.

We assume the bilayer structure generates an in-plane SH polarization orientating to right. In the trilayer ReS$_2$  shown in  Figure~\ref{fig:stru}(d), the coupling between the first and second layers should yield an in-plane SH polarization orientating to right as well. However, the in-plane SH polarization due to the coupling between the second and third layers has a reversed direction. The cancellation of the two SH polarizations results in the undetectable SH radiation. In addition, the first and third layers may couple with each other as well. However, they are symmetrical around the inversion center indicated by the arrow in  Figure~\ref{fig:stru}(d), which therefore would not generate SH polarization. However, as indicated in  Figure~\ref{fig:layer}(c), SHGs in other odd numbers of ReS$_2$ layers, such as five-layer and seven-layer, have nonzero values. It can be attributed to the small but finite optical phase shift between different SH polarizations generated by units of each adjacent layers~\cite{Li2013}, which hinders their complete cancellations with respect to the inversion center. 

The SH intensity from the tetralayer ReS$_2$ could be analyzed in the same model, as shown in  Figure~\ref{fig:layer}(e). If we consider the SH polarizations generated by second-third layers coupling and third-fourth layers coupling cancel out each other, the coupling between the first and second layers would yield an extra SH polarization. In addition, if the interlayer coupling between the first and fourth layers were considered as well, there would be  a SH polarization as well, which is smaller than that generated by the adjacent layers due to the weaker interlayer coupling. Thence, corresponding to the results shown in  Figure~\ref{fig:layer}(c), SHG in tetralayer is higher than that in bilayer. For thicker even number of ReS$_2$ layers, SHG increases gradually because of more extra SH polarizations. This increment of SH intensities in different even numbers of ReS$_2$ layers is unique compared to the variations of SH intensities in group VI TMDCs or boron nitride, which have constant net SH polarizations in different odd numbers of layers~\cite{Li2013}. The revealed SHG attribute also indicates while ReS$_2$ has the Peierls distortion, its interlayer coupling is strong enough to provide remarkable second-order nonlinearity. 

Polarization-dependent SHG is a useful technique to investigate the crystal symmetry of 2D materials. We measure the polarization dependences of the SH signals from the even numbers of ReS$_2$ layers.  In the measurements, the pump laser is converted into different linear polarizations by passing a circularly polarized laser through a rotated polarizer. Another polarizer is placed in the signal collection path, whose direction is rotated accordingly to detect SH radiation components parallel or perpendicular to the polarization of the pump laser. Figure~\ref{fig:stru}(f) displays the polar plots of the parallel and perpendicular components of the SH intensities from the six-layer ReS$_2$ shown in  Figure~\ref{fig:layer}(a). In  Figure~\ref{fig:layer}(a), a coordinate is assigned, where the $y$ axis is along the ReS$_2$'s $b$ axis and the $x$ axis is perpendicular to the $y$ axis.  The polar angle $\theta$ is the one between the laser polarization direction and $x$ axis. For the parallel and perpendicular components of the SH intensity, both of them show polarization dependences with shapes of dumbbell, which are consistent with the predictions from the low crystal symmetry in $C_1$ space group. In the parallel component, the dumbbell is twisted due to the distorted crystal structure. Also, when the polarization of the incident light is parallel to the $b$-axis, the SH signal is exactly maximum. For the perpendicular component, the polarization-dependence presents a dumbbell as well, while there is no observable distortion. The maximum SH intensity is obtained when the laser polarization is along the $x$-axis. Since the anisotropic properties, the maximum SH intensities of the parallel and perpendicular components have a ratio of $\sim$2, which yields an anisotropic distribution of the total SH intensity. This is different from that obtained from a group VI TMDC, which has an isotropic total SHG intensity~\cite{Li2013}. We also measure the polarization dependences of SH intensities from samples with different layers, as shown in  Figure~\ref{fig:stru}(g). The parallel components of SH intensities from the tetralayer and six-layer in  Figure~\ref{fig:layer}(a) present the same twisted dumbbell shape with different maxima, implying the same crystal orientation for the two flakes. These SHG's polarization dependences provide a straightforward optical method for determining the crystallographic orientation of 1T$^\prime$ ReS$_2$.

The polarization dependent SH intensities could be analyzed from the second-order nonlinear susceptibility tensor $\mathbf{d}$. We describe the SH polarizations $\mathbf{P}$ from the matrix calculation of $\mathbf{P}=\mathbf{d}\mathbf{E}$ in the $x$-$y$ coordinate defined in  Figure~\ref{fig:layer}(a). The pump laser is incident along the $z$ axis. By expressing $\mathbf{P}$ and $\mathbf{E}$ into their components in $x, y, z$ directions, the matrix calculation is given by~\cite{Shen,Boyd}

\vspace{12pt}
$\begin{bmatrix}
      P_x(2\omega)\\
      P_y(2\omega) \\
      P_z(2\omega)
           \end{bmatrix}$=$\begin{bmatrix}
d_{11} & d_{12} &d_{13} &d_{14} &d_{15} &d_{16} \\
d_{21} & d_{22} &d_{23} &d_{24} &d_{25} &d_{26}\\
d_{31} & d_{32} &d_{33} &d_{34} &d_{35} &d_{36}
           \end{bmatrix}$$\begin{bmatrix}
      E_x (\omega)E_x (\omega)\\
      E_y (\omega)E_y (\omega)\\
      E_z (\omega)E_z (\omega)\\
            2E_y (\omega)E_z (\omega)\\
            2E_x (\omega)E_z (\omega)\\
            2E_x (\omega)E_y (\omega)\\
           \end{bmatrix}$.
\vspace{12pt}

 For the electrical field components ($E_x$, $E_y$, $E_z$) at the focus plane, because of the low numerical aperture of the used objective lens, $E_z$ could be neglected. Considering an electrical field $E_0$ of the pump laser polarized along a direction with an angle $\theta$ to the $x$ axis, there are expressions of ${E_x}=E_0{\cos(\theta)}$ and ${E_y}=E_0{\sin(\theta)}$. The generated SH polarization components ($P_x$, $P_y$, $P_z$) can be regarded as electrical dipoles oscillating at the SH frequency. The SH signal radiated from $P_z$ can not be collected by the objective lens with low numerical aperture. Therefore, only $P_x$ and $P_y$ contribute to the detected  SH intensity. The parallel and perpendicular components of the SH intensities could be described as 
 
  \begin{equation}
 \begin{split}
 I_{\|}\propto(P_x\cos(\theta)+P_y\sin(\theta))^2=[d_{11}\cos^3(\theta)
+(d_{12}+2d_{26})\cos(\theta)\sin^2(\theta)\\+ (d_{21}+2d_{16})\cos^2(\theta)\sin(\theta)+d_{22}\sin^3(\theta)]^2
\end{split}
\end{equation}

 \begin{equation}
\begin{split}
 I_{\bot}\propto(P_x\sin(\theta)-P_y\cos(\theta))^2=[d_{12}\sin^3(\theta)
+(d_{11}-2d_{26})\cos^2(\theta)\sin(\theta)\\+ (2d_{16}-d_{22})\cos(\theta)\sin^2(\theta)-d_{21}\cos^3(\theta)]^2
 \end{split}
\end{equation}

 By exploiting Eqs. (1) and (2) to fit the polarization dependence in  Figure~\ref{fig:layer}(f), we could obtain the relative values of susceptibility elements of $d_{11}=0.65$, $d_{12} = 0.6$, $d_{16} =0.06$, $d_{21} =0.05$, $d_{22} = 0.05$, and $d_{26} = 0.125$. Unlike 2H TMDCs, the even-layer ReS$_2$ belongs to $C_1$ space group, the various tensor elements do not have to be equal or zero due to the low symmetry. The zero $E_z$ and uncollected $P_z$ unfortunately lead to the failed calculation of other elements, which could be measured with an assistance of a tilted far-field microscopy ~\cite{Janus1} or near-field plasmonic modes~\cite{Markus1,Park1} to extract the out-of-plane SH polarizations. 

The SHG efficiency in the bilayer ReS$_2$ is evaluated as well. We first compare it with the SHG from a chemical vapor deposition (CVD) grown monolayer WS$_2$.  With the same conditions of laser excitation and signal collection as well as the same pump power, we measure SHGs from the bilayer ReS$_2$ and monolayer WS$_2$. Figure~\ref{fig:shg}(a) plots the obtained SH spectra from the two samples when pumped by the 1558 nm pulsed laser with the same power. In the right insets, we also display the polarization-dependences of their parallel components. Anisotropic and isotropic distributions are obtained from the bilayer ReS$_2$ and monolayer WS$_2$, respectively. Note that, since the complex dispersion and nonlinear processes in the fiber laser, the output spectrum of the pump laser has multiple peaks. Correspondingly, there are multiple peaks in the SH spectra, locating between the range of 775 nm and 779 nm. While monolayer WS$_2$  is reported to have the highest second-order nonlinearity among the TMDCs~\cite{Crespi}, the SHG from the bilayer ReS$_2$ is about 8 times stronger than that from the monolayer WS$_2$. The significant SHG efficiency in the bilayer ReS$_2$ is originated from the resonant enhancementvia the excitonic effect. Figure~\ref{fig:shg}(b) displays the measured photoluminescence spectrum from the bilayer ReS$_2$, presenting exciton emission peaks at the wavelengths of 770 nm and 790 nm, which agree with the results obtained in Ref. [17]. The SH signal is on-resonant with ReS$_2$'s exciton considering their consistent photon energy, as indicated in the inset of Fig. 4(b). Due to exciton's high density of state, at the corresponding transition energy, the light-matter interaction is strongly enhanced compared to the transitions in the continuum of unbound electrons and holes,~\cite{Wang} which therefore enables a remarkable SHG response.~\cite{Malard2013,Le2016,Crespi} For the monolayer  WS$_2$ with an exciton emission around 650 nm, the off-resonance SHG has low conversion efficiency. We also pump the two samples with a femtosecond laser centered at 810 nm with a repetition rate of 80 MHz, and the obtained SH spectra are shown in Figure~\ref{fig:shg}(c). The two atomically thin materials present the comparable off-resonance SHG. 

 \begin{figure}[th!]\centering
\includegraphics[width=6in]{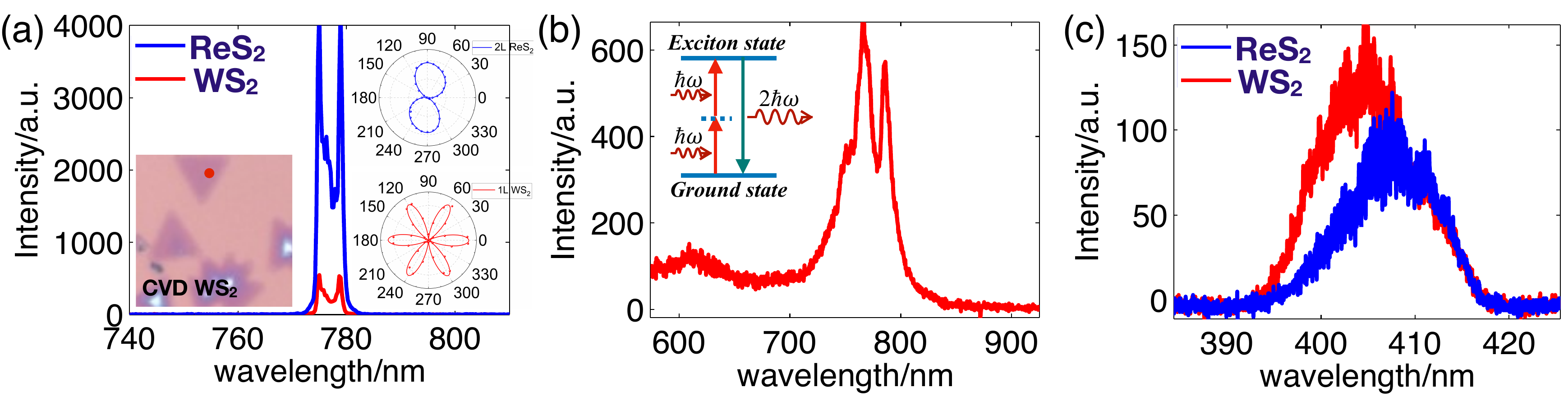}
\caption{{\small  (a) SH spectra from the bilayer ReS$_2$ and the monolayer WS$_2$ pumped by a laser at 1558 nm. Left inset: optical microscope image of the CVD monolayer WS$_2$; Right inset: polar plots of  SHG parallel components from the bilayer ReS$_2$ and monolayer WS$_2$.  (b) Photoluminescence spectrum of the bilayer ReS$_2$ pumped by a laser at 532 nm. Inset indicates the two-photon resonant SHG via excitonic effects due to the consistent photon energy of the SH signal and ReS$_2$'s exciton emission. (c) SH spectra from the bilayer ReS$_2$ and the monolayer WS$_2$ pumped by a laser at 810 nm.  }}
\label{fig:shg} 
\end{figure}

The above experiment results indicate the bilayer ReS$_2$ has a high SHG efficiency with a pump around 1558 nm due to the two-photon resonance with its exciton. We then calculate the absolute second-order sheet susceptibility at this wavelength by normalizing the SH intensity from ReS$_2$ ($I_{ReS_2}$) to the reference SH intensity from the surface of a $z$-cut bulk crystal of lithium niobate ($I_{LN}$) when pumped by the 1558 nm laser, which can be expressed as~\cite{Clark} 

\begin{equation}
\chi_{ReS_2}^{(2)}=\frac{1}{16\pi\Delta{k_{LN}}\Delta{h}}\frac{[n_{LN}(\omega)+1]^3}{n_{LN}(\omega)n_{LN}^{1/2}(2\omega)}\left(\frac{I_{ReS_2}(2\omega)}{I_{LN}(2\omega)}\right)^{1/2}\chi_{LN}^{(2)}
\end{equation}

where $\Delta{h}\sim1.6$ nm is the thickness of the bilayer ReS$_2$, $n_{LN}(\omega)$ and $n_{LN}(2\omega)$  are refractive indices of the light with frequencies of $\omega$ and $2\omega$ in lithium niobate, which define their wavenumber difference $\Delta{k}_{LN}$, and $\chi_{LN}^{(2)}$ is lithium niobate's nonlinear  susceptibility. By counting $\Delta{k}_{LN}$ and  $\chi_{LN}^{(2)}$, $\chi_{ReS_2}^{(2)}$ is estimated as 900 pm/V. Because of resonant enhancement via the excitonic effect, the bilayer ReS$_2$ has an effective nonlinear susceptibility that is remarkably large among the 2D materials when pumped by a telecom-band laser.


In summary, we experimentally observed unexpected layer-dependent, strong, and anisotropic SHGs in atomically thin ReS$_2$. Different from the well-studied group VI TMDCs, ReS$_2$ as the group VII TMDC has a distorted CdCl$_2$ crystal structure. While its monolayer, other odd numbers of layers and bulk crystal belong to $C_i$ space group and have inversion symmetry, the inversion symmetries in even numbers of layers are broken. Thus, ReS$_2$ presents strong (negligible) SHG in even (odd)  layer thickness, which is opposite to layer-dependence of SHG in group VI TMDCs. The exactly zero  SHG in ReS$_2$ monolayer and strong SHGs in even numbers of ReS$_2$ layers indicate the unexpected strong interlayer coupling, which generated appreciable SH polarizations.  In a bilayer ReS$_2$, we obtain an effective second-order nonlinear susceptibility of 900 pm/V pumped by a telecom-band laser, which is remarkably high among those reported in 2D materials.  The laser polarization dependence of ReS$_2$'s SHG is strongly anisotropic and  indicates its distorted lattice structure with more unequal and non-zero second-order susceptibility elements.  Our results not only reveal another distinct property between group VI and group VII TMDCs, but also open up opportunities for studying piezoelectricity and valley-dependent physics  in even numbers of ReS$_2$ layers arising from the broken inversion symmetry.


\begin{acknowledgement}

 Financial support was provided by the Key Research and Development Program (2017YFA0303800, 2016YFA0301204), NSFC (61522507, 61775183, 11634010, 11474277 and 11434010),  the Key Research and Development Program in Shaanxi Province of China (2017KJXX-12), and the Fundamental Research Funds for the Central Universities (3102017jc01001, 3102018jcc034). We thank Wei Ji and Jingsi Qiao for many fruitful discussions as well as the AFM measurements by the Analytical \& Testing Center of NPU. 

\end{acknowledgement}



\begin{thebibliography}{99}

\bibitem{Fai2013}Mak, K. F.; He, K.; Lee, C.; Lee, G. H.; Hone, J.; Heinz, T. F.; Shan, J. Tightly Bound Trions in Monolayer MoS$_2$. $\it{Nat.~Mater.}$  $\bf{2013}$, 12, 207-211.
\bibitem{Shan2016}Lee, J.; Mak, K. F.; Shan, J. Electrical Control of the Valley Hall Effect in Bilayer MoS$_2$ Transistors. $\it{Nat.~Nanotechnol.}$  $\bf{2015}$, 11, 421-425.
\bibitem{Xu2017}Ross, J. S.; Rivera, P.; Schaibley, J.; Lee-Wong, E.; Yu, H.; Taniguchi, T.; Watanabe, K.; Yan, J.; Mandrus, D.; Cobden, D.; Yao, W.; Xu, X. Interlayer Exciton Optoelectronics in a 2D Heterostructure p-n Junction. $\it{Nano~Lett.}$  $\bf{2017}$, 17, 638-643. 
\bibitem{LiaoAM}Wang, J.; Yao, Q.; Huang, C.; Zou, X.; Liao, L.; Chen, S.; Fan, Z.; Zhang, K.; Wu, W. W.; Xiao, X.; Jiang, C.; Wu, W. High Mobility MoS$_2$ Transistor with Low Schottky Barrier Contact by Using Atomic Thick h-BN as a Tunneling Layer. $\it{Adv.~Mater.}$  $\bf{2016}$, 28, 8302-8308.
\bibitem{Xinran}Yu, Z.; Ong, Z. Y.; Li, S.; Xu, J. B.; Zhang, G.; Zhang, Y. W.; Shi, Y.; Wang, X. Analyzing the Carrier Mobility in Transition-Metal Dichalcogenide MoS$_2$ Field-Effect Transistors. $\it{Adv.~Funct.~Mater.}$  $\bf{2017}$, 27.
\bibitem{JianluAM}Wang, X.; Wang, P.; Wang, J.; Hu, W.; Zhou, X.; Guo, N.; Huang, H.; Sun, S.; Shen, H.; Lin, T.; Tang, M.; Liao, L.; Jiang, A.; Sun, J.; Meng, X.; Chen, X.; Lu, W.; Chu, J. Ultrasensitive and Broadband MoS$_2$ Photodetector Driven by Ferroelectrics. $\it{Adv.~Mater. }$  $\bf{2015}$, 27, 6575-6581.
\bibitem{Xiaodong}Wu, S.; Buckley, S.; Schaibley, J. R.; Feng, L.; Yan, J.; Mandrus, D.; Hatami, F.; Yao, W. M.; Vuckovic, J.; Majumdar, A.; Xu, X. Monolayer Semiconductor Nanocavity Lasers with Ultralow Thresholds. $\it{Nature}$  $\bf{2015}$, 520, 69-72. 
\bibitem{Yuye}Ye, Y.; Wong, Z. J.; Lu, X.; Ni, X.; Zhu, H.; Chen, X.; Wang, Y,; Zhang, X. Monolayer Excitonic Laser. $\it{Nat.~Photon. }$  $\bf{2015}$, 9, 733.
\bibitem{Friemelt}Friemelt, K.; Luxsteiner, M. Ch.; Bucher, E. Optical Properties of The Layered Transition-Metal-Dichalcogenide ReS$_2$: Anisotropy in the Van Der Waals Plane. $\it{ J.~Appl.~Phys. }$  $\bf{1993}$, 74, 5266.
\bibitem{Lin}Lin, Y.; Komsa, H.; Yeh, C. H.; Bjorkman, T.; Liang, Z. Y.; Ho, C. H.; Huang, Y. S.; Chiu, P.; Krasheninnikov, A. V.; Suenaga, K. Single-Layer ReS$_2$: Two-Dimensional Semiconductor with Tunable In-Plane Anisotropy. $\it{ACS~Nano}$  $\bf{2015}$, 9, 11249-11257.
\bibitem{Hone}Chenet, D.; Aslan, O. B.; Huang, P. Y.; Fan, C.; Der Zande, A. M.; Heinz, T. F.; Hone, J. In-Plane Anisotropy in Mono- and Few-Layer ReS$_2$ Probed by Raman Spectroscopy and Scanning Transmission Electron Microscopy. $\it{Nano~Lett. }$  $\bf{2015}$, 15, 5667-5672.
\bibitem{Etienne}Lorchat, E.; Froehlicher, G.; Berciaud, S. Splitting of Interlayer Shear Modes and Photon Energy Dependent Anisotropic Raman Response in N-Layer ReSe$_2$ and ReS2$_2$. $\it{ACS~Nano}$  $\bf{2015}$, 10, 2752-2760.
\bibitem{MiaoNC}Liu, E.; Fu, Y.; Wang, Y.; Feng, Y.; Liu, H.; Wan, X.; Zhou, W.; Wang, B.; Shao, L.; Ho, C.; Huang, Y.; Cao, Z.; Wang, L.; Li, A.; Zeng, J.; Song, F.; Wang, X.; Shi, Y.; Yuan, H.; Hwang, H. Y.; Cui, Y.; Miao, F.; Xing, D. Y. Integrated Digital Inverters Based on Two-dimensional Anisotropic ReS$_2$ Field-effect Transistors. $\it{Nat.~Commun.}$  $\bf{2015}$, 6, 6991-6991.
\bibitem{zhaohui}Cui, Q.; He, J.; Bellus, M. Z.; Mirzokarimov, M.; Hofmann, T.; Chiu, H.; Antonik, M.; He, D.; Wang, Y.; Zhao, H. Transient Absorption Measurements on Anisotropic Monolayer ReS$_2$. $\it{Small}$  $\bf{2015}$, 11, 5565-5571.
\bibitem{Wolverson}Wolverson, D.; Crampin, S.; Kazemi, A. S.; Ilie, A.; Bending, S. J. Raman Spectra of Monolayer, Few-Layer, and Bulk ReSe$_2$: An Anisotropic Layered Semiconductor. $\it{ACS~Nano}$  $\bf{2014}$, 8, 11154-11164.
\bibitem{Weiji}Qiao, X.; Wu, J.; Zhou, L.; Qiao, J.; Shi, W.; Chen, T.; Zhang, J.; Ji, W.; Tan, P. Polytypism and Unexpected Strong Interlayer Coupling of two-Dimensional Layered ReS$_2$. $\it{Nanoscale}$  $\bf{2016}$, 8, 8324-8332.
\bibitem{Heinz}Aslan, O. B.; Chenet, D.; Der Zande, A. M.; Hone, J.; Heinz, T. F. Linearly Polarized Excitons in Single- and Few-Layer ReS$_2$ Crystals. $\it{ACS~Photon.}$  $\bf{2016}$, 3, 96-101.
\bibitem{Arora}Arora, A.; Noky, J.; DrŸppel, M.; Jariwala, B.; Deilmann, T.; Schneider, R.; Schmidt, R.; Pozo-Zamudio, O. D.; Stiehm, T.; Bhattacharya, A.; Krger, P.; Vasconcellos, S. M. D.; Rohlfing, M.; Bratschitsch, R. Highly Anisotropic in-Plane Excitons in Atomically Thin and Bulklike 1 T$^\prime$-ReSe$_2$. $\it{Nano~Lett.}$  $\bf{2017}$, 17, 3202-3207.
\bibitem{FaxianACS}Zhang, E.; Wang, P.; Li, Z.; Wang, H.; Song, C.; Huang, C.; Chen, Z.; Yang, L.; Zhang, K.; Lu, S.; Wang, W.; Liu, S.; Fang, H.; Zhou, X.; Yan, H.; Zou, J.; Wan, X.; Zhou, P.; Hu, W.; Xiu, F. Tunable Ambipolar Polarization-Sensitive Photodetectors Based on High-Anisotropy ReSe$_2$ Nanosheets. $\it{ACS~Nano}$  $\bf{2016}$, 10, 8067-8077.
\bibitem{Tongay}Tongay, S.; Sahin, H.; Ko, C.; Luce, A.; Fan, W.; Liu, K.; Zhou, J.; Huang, Y. S.; Ho, C. H.; Yan, J.; Ogletree, D. F.; Aloni, S.; Ji, J.; Li, S.; Li, J.; Peeters, F. M.; Wu, J. Monolayer Behavior in Bulk ReS$_2$ Due to Electronic and Vibrational Decoupling. $\it{Nat.~Commun.}$  $\bf{2013}$, 5, 1-6.
\bibitem{FaxianAFM}Zhang, E.; Jin, Y.; Yuan, X.; Wang, W.; Zhang, C.; Tang, L.; Liu, S.; Zhou, P.; Hu, W.; Xiu, F. ReS$_2$-Based Field-Effect Transistors and Photodetectors. $\it{Adv.~Funct.~Mater.}$  $\bf{2015}$, 25, 4076-4082.
\bibitem{Malard2013}Malard, L. M.; Alencar, T. V.; Barboza, A. P. M.; Mak, K. F.;  Paula, A. M. D. Observation of Intense Second Harmonic Generation from MoS$_2$ Atomic Crystals. $\it{Phy.~Rev.~B}$  $\bf{2013}$, 87. 
\bibitem{Li2013}Li, Y.; Rao, Y.; Mak, K. F.; You, Y.; Wang, S.; Dean, C. R.; Heinz, T. F. Probing Symmetry Properties of Few-Layer MoS$_2$ and h-BN by Optical Second-Harmonic Generation. $\it{Nano~Lett.}$  $\bf{2013}$, 13, 3329.
\bibitem{Tan2011}Tan, P.H.; Han, W.P.; Zhao, W. J.; Wu, Z.H.; Chang, K.; Wang, H.; Wang, Y.F.; Bonini, N.; Marzari, N.; Pugno, N.; Savini, G.; Lombardo, A.; Ferrari, A. C. The Shear Mode of Multilayer Graphene. $\it{Nat.~Mater.}$  $\bf{2012}$, 11, 294.
\bibitem{Liang2017} Liang, L.; Zhang, J.; Sumpter, B. G.; Tan, Q. H.; Tan, P. H.; Meunier, V. Low-Frequency Shear and Layer-Breathing Modes in Raman Scattering of Two-Dimensional Materials. $\it{ACS~ Nano}$  $\bf{2017}$, 11, 11777. 
\bibitem{Le2016}Le, C. T.; Clark, D. J.; Ullah, F.; Senthilkumar, V.; Jang, J. I.; Sim, Y.; Seong, M.-J.; Chung, K.-H.; Park, H.; Kim, Y. S., Nonlinear Optical Characteristics of Monolayer MoSe$_2$. $\it{Annalen~der~Physik}$  $\bf{2016}$, 528 (7-8), 551-559.
\bibitem{Zhang2015}Zhang, S.; Dong, N.; McEvoy, N.; O'Brien, M.; Winters, S.; Berner, N. C.; Yim, C.; Li, Y.; Zhang, X.; Chen, Z.; Zhang, L.; Duesberg, G. S.; Wang, J., Direct Observation of Degenerate Two-Photon Absorption and Its Saturation in WS$_2$ and MoS$_2$ Monolayer and Few-Layer Films. $\it{ ACS~Nano}$  $\bf{2015}$, 9 (7), 7142-7150.
\bibitem{Janus1} Lu, A. Y.; Zhu, H.; Xiao, J.; Chuu, C. P.; Han, Y.; Chiu, M. H.; Cheng, C. C.; Yang, C. W.; Wei, K. H.; Yang, Y.; Wang, Y.; Sokaras, D.; Nordlund, D.; Yang, P.; Muller, D.; Chou, M. Y.; Zhang, X.;  Li, L. J. Janus Monolayers of Transition Metal Dichalcogenides. $\it{Nat.~Nanotechnol.}$  $\bf{2017}$, 12, 744-749.
\bibitem{Markus1} Park, K. D.; Jiang, T.; Clark, G.; Xu, X.;Raschke, M. B. Radiative Control of Dark Excitons at Room Temperature by Nano-Optical Antenna-Tip Purcell Effect. $\it{Nat.~Nanotechnol.}$  $\bf{2018}$, 13, 59-64.
\bibitem{Park1} Zhou, Y.; Scuri, G.; Wild, D. S.; High A. A.; Dibos, A.; Jauregui, L. A.; Shu, C.; Greve, K. D.; Pistunova, K.; Joe, A. Y.; Taniguchi, T.; Watanabe, K.; Kim, P.; Lukin, M. D.; Park, H. Probing Dark Excitons in Atomically Thin Semiconductors via Near-Field Coupling to Surface
plasmon polaritons. $\it{Nat.~Nanotechnol.}$  $\bf{2017}$, 12, 856-860.
\bibitem{Markus2} Kravtsov, V.; AlMutairi, S.; Ulbricht, R.; Kutayiah, A. R.; Belyanin, A.; Raschke, M. B., Enhanced Third-Order Optical Nonlinearity Driven by Surface-Plasmon Field Gradients. $\it{ Phys. Rev. Lett.}$  $\bf{2018}$, 120 (20), 203903.
\bibitem{Shen}Shen, Y. R. The Principles of Nonlinear Optics. $\it{Wiley-Interscience: New York}$  $\bf{2003}$.
\bibitem{Boyd}Boyd, R. W. Nonlinear Optics, 3rd ed. $\it{Academy~ Press: San~Diego, CA}$  $\bf{2008}$.
\bibitem{Crespi}Janisch, C.; Wang, Y.; Ma, D.; Mehta, N.; Elias, A. L.; Perealopez, N.; Terrones, M.; Crespi, V. H.; Liu, Z. Extraordinary Second Harmonic Generation in Tungsten Disulfide Monolayers. $\it{Sci.~Rep.}$  $\bf{2014}$, 4, 5530
\bibitem{Wang}Wang, G.; Chernikov, A.; Glazov, M. M.; Heinz, T. F.; Marie, X.; Amand, T.; Urbaszek, B., Colloquium
: Excitons in Atomically Thin Transition Metal Dichalcogenides.  $\it{Rev.~Mod.~Phys.}$  $\bf{2018}$, 90, 021001.
\bibitem{Clark}Clark, D. J.; Senthilkumar, V.; Le, C. T.; Weerawarne, D. L.; Shim, B.; Jang, J. I.; Shim, J. H.; Cho, J.; Sim, Y.; Seong, M. J.; Rhim, S. H.; Freeman, A. J.; Chung, K. H.; Kim, Y. S. Strong Optical Nonlinearity of CVD-Grown MoS$_2$ Monolayer as Probed by Wavelength-Dependent Second-Harmonic Generation. $\it{Phy.~Rev.~B}$  $\bf{2014}$, 90, 121409.

\end{thebibliography}

\end{document}